# Design & development of position sensitive detector for hard X-ray using SiPM and new generation scintillators

S. K. Goyal*[a, b], Amisha P. Naik[a], Mithun N. P. S.[b], S. V. Vadawale[b], Neeraj K. Tiwari[b], T. Chattopadhyay[b,c], N. Nagrani[b], S. Madhavi[b], T. Ladiya[b], A. R. Patel[b], M. Shanmugam[b], H. L. Adalja[b], V. R. Patel[b], G. P. Ubale[b]

[a]Institute of Technology, Nirma University, Chandlodia, Ahmedabad, Gujarat, INDIA 382481; [b]Physical Research Laboratory, Navrangpura, Ahmedabad, Gujarat, INDIA 380009; [c]Pennsylvania State University, Old Main, State College, PA 16801, USA

## ABSTRACT

There is growing interest in high-energy astrophysics community for the development of sensitive instruments in the hard X-ray energy extending to few hundred keV. This requires position sensitive detector modules with high efficiency in the hard X-ray energy range. Here, we present development of a detector module, which consists of 25 mm x 25 mm $CeBr_3$ scintillation detector, read out by a custom designed two dimensional array of Silicon Photo-Multipliers (SiPM). Readout of common cathode of SiPMs provides the spectral measurement whereas the readout of individual SiPM anodes provides measurement of interaction position in the crystal. Preliminary results for spectral and position measurements with the detector module are presented here.

## 1. INTRODUCTION

Sensitive spectral, timing and polarimetric observations in hard X-ray energy range extending to few 100 keV is important in addressing a wide range of scientific problems like the jet contributions to hard X-ray spectrum of black hole binaries, prompt emission mechanism in Gamma Ray Bursts (GRBs), emission mechanism and geometry in pulsars etc. However, there is an order of magnitude difference in sensitivity of instruments beyond 80 keV, compared to those at lower energies. Development of hard X-ray telescopes like NuSTAR[1] enhanced the sensitivity in the energy range of 20-80 keV, but the current hard X-ray optics technology is limited to 80 keV. Attempts are being made to build X-ray telescopes working at energies up to 200 keV[2]. Such experiments would require position sensitive focal plan detectors to be used along with the optics, having significant efficiency till several hundred keV. An alternative to focusing telescopes in high energies is large area Compton telescope. Development of Compton telescopes also would require position sensitive hard X-ray detector modules. In this context, we are developing position sensitive hard X-ray detector module using new generation scintillators and SiPMs having sub-pixel spatial resolution and high detection efficiency in the energy range of 10-400 keV.

### 1.1 Position sensitive hard X-ray detector module – design goals

Primary requirement of the detector module is to have high detection efficiency in the energy range of 10-400 keV with high spatial resolution. The detector module design should be compact so that it can be used as focal plane detector as well as it is feasible to realize large area arrays. Measurement of interaction position requires readout of multiple channels (pixels) of the detector module; hence, sub-pixel resolution would offer better position sensitivity with lesser number of readout channels. Low operating voltage can also be one of the desired specifications while designing high-energy detector modules as most of the detectors in this energy range require high operating voltages (hundreds of Volts).

Semiconductor detectors such as CZT / CdTe are typically used for such application. However, in these detectors, achieving high spatial resolution for thick detectors (with good efficiency at high energies) is difficult and sub-pixel sampling is not possible. The complicated spectral response, anomalous pixel behavior and requirement of high operating voltages are other undesirable characteristics of these semiconductor detectors.

An alternative is hard X-ray detector module of scintillator readout by array of SiPM. In this work, we present the development of prototype detector module with new generation scintillator and custom array of SiPMs along with its preliminary characterization.

*goyal@prl.res.in; phone 91 79 26314932; www.prl.res.in

# 2. DEVELOPMENT OF SCINTILLATOR-SIPM DETECTOR MODULE

Hard X-ray imaging detector is designed using scintillation detector and custom designed array of SiPM. We have carried out the experiment using $CeBr_3$ scintillation detector with 6x6 2D array of SiPM (MicroC series 30035 – SensL make). The experiment is also carried out using CsI(Tl) detector with 4x4 2D array of SiPM (MicroC series 30035). The front-end electronics (FEE) has been designed for the readout of SiPM and tuned as per the scintillation gain. The shaping output of the common cathode is used for the measurement of the energy of the incident X-rays and as the trigger for the readout of anodes.

In our earlier study[3], we used CsI (Tl) scintillation detector along with 4x4 SensL make SiPM array, for the proof of concept for the detector module. Here we have used $CeBr_3$ scintillation detector because of its higher light output and faster response.

## 2.1 Scintillation detector

There are many scintillation crystals, which can be used for the detection of X-rays. Table 1 provides a comparison between some of the crystals.

Table 1. Comparison table for various scintillation detectors.

|  | $LaBr_3$:Ce | NaI (Tl) | CsI (Tl) | $CeBr_3$ |
|---|---|---|---|---|
| Density (g/cm$^3$) | 5.1 | 3.7 | 4.5 | 4.5 |
| Light photons (photons per keV) | 63 | 38 | 54 | 60 |
| FWHM at 662 keV | 3% | 8% | 8% | 3.8% |
| Decay constant (ns) | 16 | 250 | 1000 | 19 |
| Peak wavelength (mm) | 380 | 415 | 550 | 380 |
| Hygroscopic | Yes | Yes | Slightly | Yes |

The newly developed $LaBr_3$:Ce and $CeBr_3$ are the new generation crystals in the family of scintillation detector and have an advantage over conventional room temperature detectors. $LaBr_3$:Ce has an internal background at 789 keV and 1436 keV, which makes it difficult to use. $CeBr_3$ has high light yield ~60 photons per keV and is having shorter decay time, which makes it suitable for the SiPM readout compared to NaI (Tl) and CsI (Tl). Since $CeBr_3$ is a hygroscopic crystal, we have used it in a hybrid packing. The active area of the crystal used is 25 mm x 25 mm having thickness of 5 mm. The total size of the hybrid pack is 33 mm x 33 mm. It has transparent quartz window on the photon readout side. Figure 1 is the photographic view of the $CeBr_3$ crystal.

We have also used CsI (Tl) crystal of size 15 mm x 15 mm having thickness of 3 mm. Figure 2 shows the photograph of the CsI (Tl) crystal. CsI (Tl) crystal is easy to handle in the lab because it is less hygroscopic. It has decay time of ~1000 ns. We used this crystal to see the effect of the SiPM noise in terms of the slow and fast scintillation decay constants.

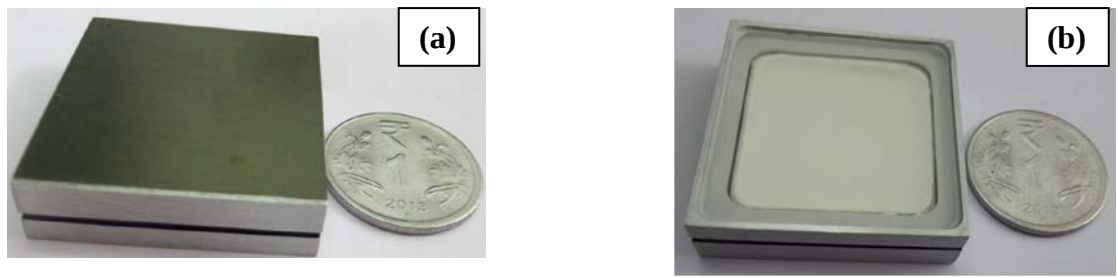

Figure 1. $CeBr_3$ crystals. (a) Top view, (b) Bottom view (photon readout side).

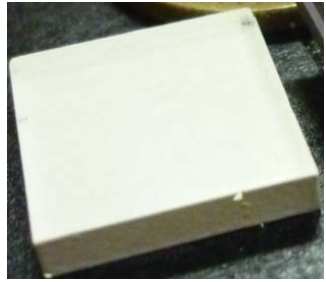

Figure 2. CsI (Tl) crystal (Top – view) of size 15 mm x 15 mm x 3 mm.

## 2.2 Silicon photomultiplier (SiPM)

SiPM is a new development in the field of photon detection and can be described as 2D array of small avalanche photodiodes (APD). SiPM is an alternate option of using conventional photo multiplier tubes (PMTs), as it offers comparable photon detection efficiency, small size, feasibility of compact array, low cost, excellent time resolution, low operating voltage and insensitivity to the external magnetic field. In SiPM, APDs work in Geiger mode. The current through APD is controlled using in-built quenching resistor. The current flow through the parallel combination of APDs is linearly proportional to the number of incident photons. We are using SensL make MicroC series SiPM (30035) for our experiment. The technical specifications of the SiPM are listed in table 2.

Table 2. Technical specifications of SiPM.

| **Parameter** | **Value** |
|---|---|
| Max breakdown voltage | 24.7 V |
| Overvoltage range | 1.0 V – 5.0 V |
| Spectral range | 300 nm – 950 nm |
| Peak wavelength ($\lambda_p$) | 420 nm |
| Photo detection efficiency (at $\lambda_p$) – $V_{br}$+2.5V | ~31% |
| Gain (anode to cathode readout) – $V_{br}$+2.5V | $3 \times 10^6$ |
| Dark current – $V_{br}$+2.5V | ~150 nA |
| Dark count rate – $V_{br}$+2.5V | 300 kHz |
| Microcell recharge time constant | 82 ns |
| Cathode (anode-cathode) – $V_{br}$+2.5V | 850 pF |
| Temperature dependence of $V_{br}$ | 21.5 mV/ºC |
| Temperature dependence of gain | -0.8 %/ºC |
| Crosstalk – $V_{br}$+2.5V | 7% |
| After-pulsing – $V_{br}$+2.5V | 0.2% |
| Active area | 3 mm x 3 mm |
| No. of microcells | 4774 |
| Microcell fill factor | 64% |
| Package dimensions | 4 mm x 4 mm |
| Operating temperature range | -40 ºC to +85 ºC |
| Encapsulant refractive index | 1.59 @ 420 nm |

Figure 3 shows the photographs of the SiPM. 6x6 2D array has been designed using this SiPM. The gap between two SiPMs is 0.2 mm, as designed in the PCB layout. The total area covered by the 36 SiPMs is 25 mm x 25 mm. Total PCB size is 33 mm x 33 mm. This 6x6 SiPM array has been used for the photons readout of CeBr$_3$ scintillator detector. Photograph shown in 3 (d) is 4x4 array of SiPM, which has been used for the readout of the

3 | P a g e

CsI (Tl) scintillator. The area covered by 16 SiPMs is 16.6 mm x 16.6 mm and the total PCB size is 20 mm x 20 mm.

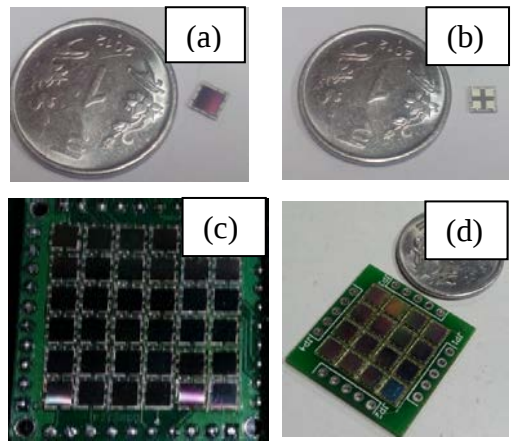

Figure 3. (a) Top view of SiPM, (b) Bottom view of SiPM, (c) 6x6 array of SiPM, (d) 4x4 array of SiPM.

The selection of the scintillator and SiPM for use in position sensitive X-ray detector involves consideration of various aspects such as:
- Pixel size, dead area, active area, dark current, operating voltage, wavelength, photo detection efficiency, cross talk etc. are the parameters for the selection of SiPM.
- SiPM becomes saturated, when the number of incident photons on the SiPM are comparable to the number of the microcells available. This limits the higher energy range of the X-ray.
- Lower energy threshold is decided based on the dark noise of the SiPM and photons output of the scintillator. Higher is the scintillator output, higher is the SNR but has the penalty on the high-energy detection (SiPM saturation).
- Faster is the scintillator decay time constant; lower is the integrated dark counts from the SiPM in the readout electronics.

## 3. EXPERIMENTAL SETUP FOR THE ENERGY MEASUREMENT

There are total 36 SiPMs in the detector module. For the energy measurement and to see the lower energy threshold, all 36 anodes are connected to ground and all 36 cathodes are biased to a common supply. The output from the common cathode has been used for the energy measurement.

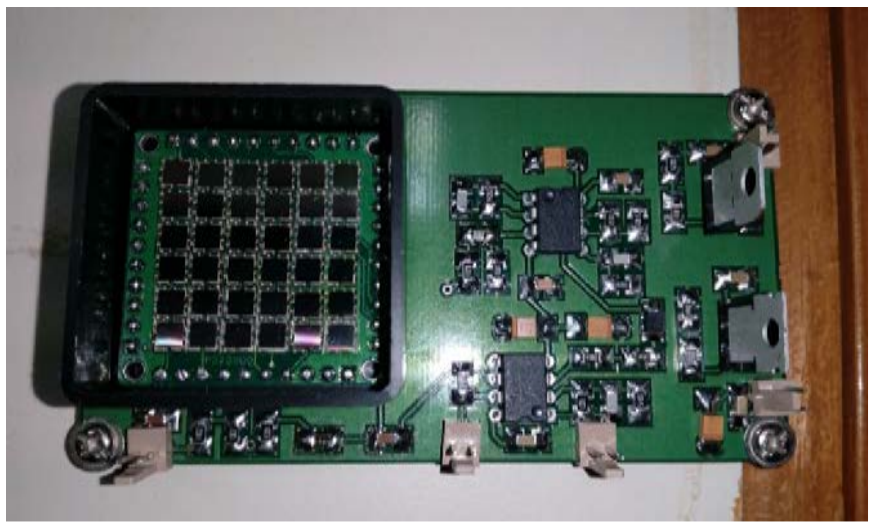

Figure 4. Front End Electronics (FEE) card for the common cathode readout.

The CeBr$_3$ scintillator and SiPM array are kept in a light tight black Delrin box. At present, there is no optical coupling used between scintillator and SiPM. The common cathode has been biased to a positive supply of +28 Vdc through a biasing resistor. This output is AC coupled to the RC feedback type charge sensitive preamplifier (CSPA). The output of the CSPA is an exponential decay pulse, where peak amplitude refers to the energy of the incident X-ray and fall time is decided by the RC time constant. The CSPA output is shaped to Gaussian

4 | P a g e

pulse using CR-RC-RC circuit where peaking time constant has been set to 0.3 µs. The shaping output has been read-out using multi-channel analyzer (EASY MCA 8K) to get the spectrum output.

### 3.1 Results for CeBr$_3$ + SiPM array (6x6)

Figure 5 shows the energy spectrum from the detector module of CeBr$_3$ and 6x6 SiPM array. $^{241}$Am and $^{109}$Cd radioactive sources are used for the experiment. Figure 5 (a) shows the spectrum of the $^{241}$Am source. This spectrum clearly shows the detection of the 14 keV + 18 keV energy lines, but these lines are merging because of the poor energy resolution. Similarly, $^{109}$Cd results are shown in the figure 5 (b), where 22 keV and 88 keV energy lines are clearly seen. Figure 5 (c) is the background data without any radioactive source.

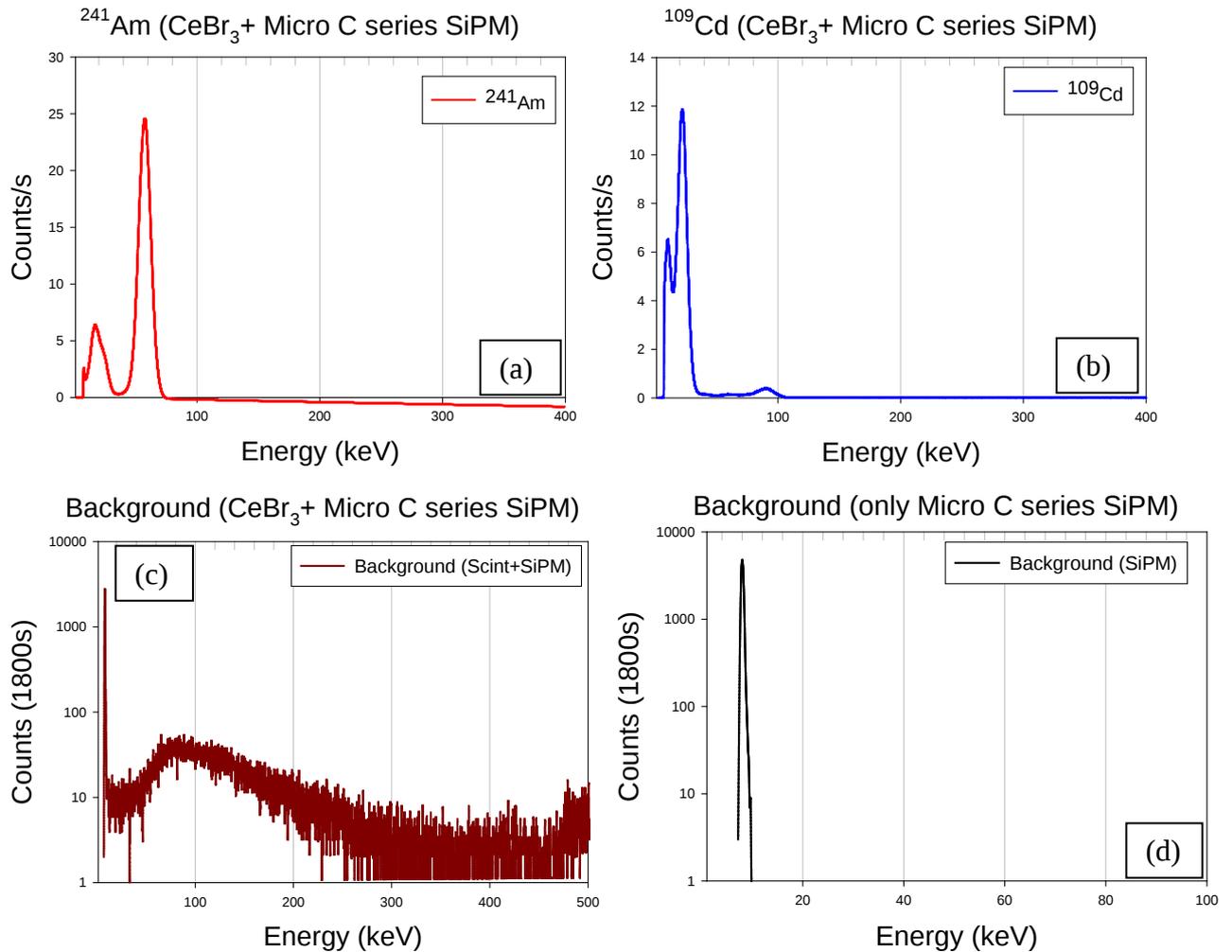

Figure 5. Preliminary results using CeBr3, and 6x6 array of SiPM – Energy measurement of the incident X-ray.

In this figure, only scintillator + SiPM readout is plotted. Figure 5 (d) shows the electronics noise of the only SiPM. In this figure, scintillator was removed and only SiPM data was plotted. All these measurements (figure 5) are carried out at pulse peaking time of 0.3 µs. This pulse peaking time is limited by the minimum rise time of the MCA. Since scintillator decay time is shorter (~19 ns), we expect better performance when we go for the lower pulse peaking time, which will make signal to noise ratio better. We are further carrying out the experiment using ASIC where we can go to different pulse peaking times such as 50 ns, 100 ns and 150 ns. The characterization of the performance with different shaping times is under progress.

### 3.2 Results for CsI (Tl) + SiPM array (4x4)

CsI (Tl) results using 4x4 2D array of SiPM are shown in figure 6. Figure 6 (a) is the result using $^{241}$Am source, 6 (b) is the result using $^{109}$Cd source. Figure 6 (c) is the background data using CsI (Tl) scintillator and SiPM array, while in figure 6 (d) is the electronic noise of the SiPM without scintillator. Pulse peaking time of 2 µs is used for the CsI (Tl) scintillator readout.

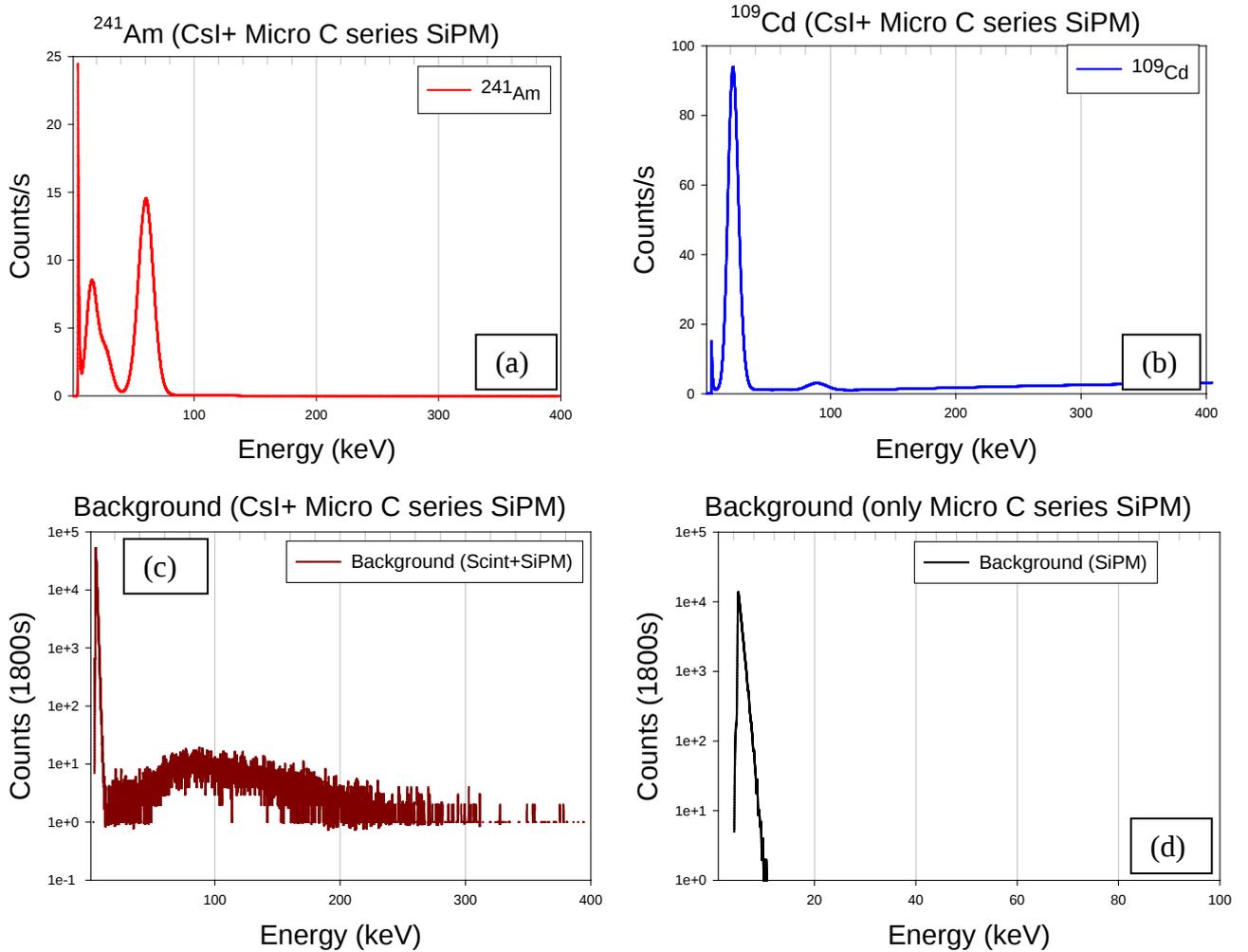

Figure 6. Preliminary results using CsI (Tl), and 4x4 array of SiPM – Energy measurement of the incident X-ray.

## 4. EXPERIMENTAL SETUP FOR HARD X-RAY POSITION DETECTION

### 4.1 Experimental setup using Vertilon

Here we present the experimental setup or interaction position measurement with CeBr$_3$ scintillator and 6x6 SiPM array. There are total 36 SiPMs, where total 36 anodes are used for the position readout and all 36 cathodes are shorted together to a common supply (28 Vdc) using a biasing resistor. We are using both Vertilon based readout and ASIC readout of all the anodes for the position measurement. For the Vertilon based readout, we have used Vertilon make data acquisition system (IQSP482), 32-channel SMB distribution system (SDS232) and control & acquisition software. The block schematic for the electrical connections between various systems is shown in figure 7.

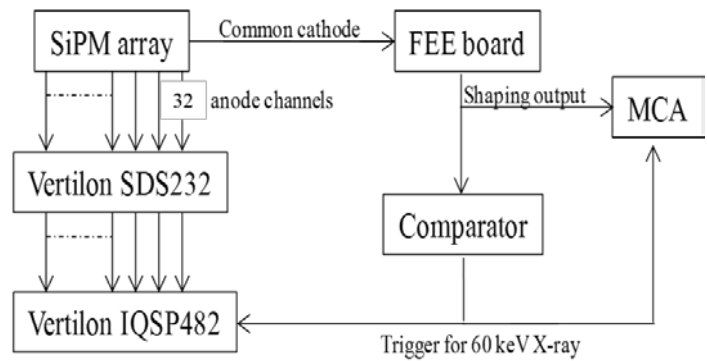

Figure 7. Block schematic for the electrical interconnections of various systems.

Here SiPM array is reading the photons output of the CeBr$_3$ scintillator (scintillator is not shown in the block schematic). Figure 8 shows the actual photograph of the experimental setup for the position measurement of the incident X-rays. Vertilon SDS232 system can have only 32 anodes as the inputs, in this case only 32 SiPMs anodes (from 36 anodes) are connected to the SDS232 (4 anodes from the 4 corners are not connected and are left open). These connections are further interfaced to the Vertilon IQSP482 system. The Vertilon data acquisition software acquires the 32 parallel data from all the channels on the common cathode trigger. In this setup, SiPM array is connected to the FEE board which has only CSPA, the output is given to another board having shaping time of 0.3 µs for the energy measurement. The trigger for the energy is generated using LM319 IC high-speed dual comparator IC. Here we have used comparator threshold sufficiently high so that only 60 keV energy line ($^{241}$Am radioactive source) is registered and hence 14 keV and 18 keV energy lines will not be counted.

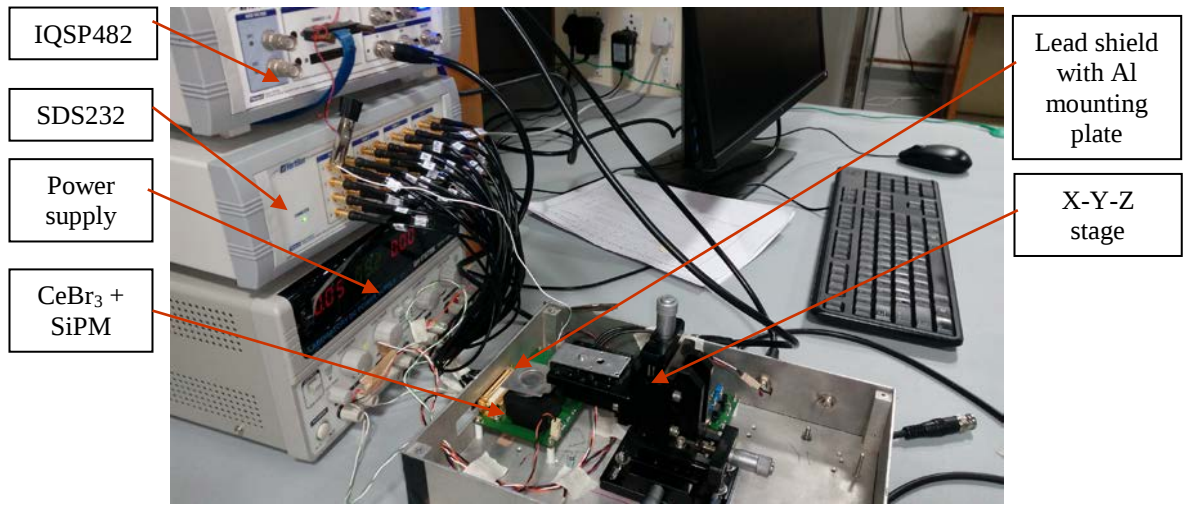

Figure 8. Experimental setup for the position and energy measurement of the X-rays.

The Vertilon setup is reading all 32 anodes in the external trigger mode with an input from the comparator output. The same comparator output is given to the MCA also for the gated trigger mode to register the energy information. We are using XYZ positioning stage for the movement of the X-ray radioactive source. The source is kept on a 5 mm thick lead shield with a central aperture of ~0.5 mm diameter. X-rays are allowed to interact with the scintillator through this hole.

**4.1.1 Results with Vertilon setup for 1 mm movement and 0.5 mm aperture**

Vertilon software provides output in terms of the charge (pC) for all 32 pixels separately. There are always some background events registered based on the dark count rate of SiPM. These events are very random in nature and can occur in any of these pixels. In our experiment, we are adding up the charge values of all the 32 pixels together for each event separately to get the energy information. Figure 9 shows the histogram for one of the cases. As we are allowing the 60 keV energy line, the events for which total charge is coming under the area of 20 pC to 35 pC are counted and rest of the events are rejected. MCA is also connected in the gated mode with inputs from the comparator and pulse shaping amplifiers, to get the energy information of the incident photon.

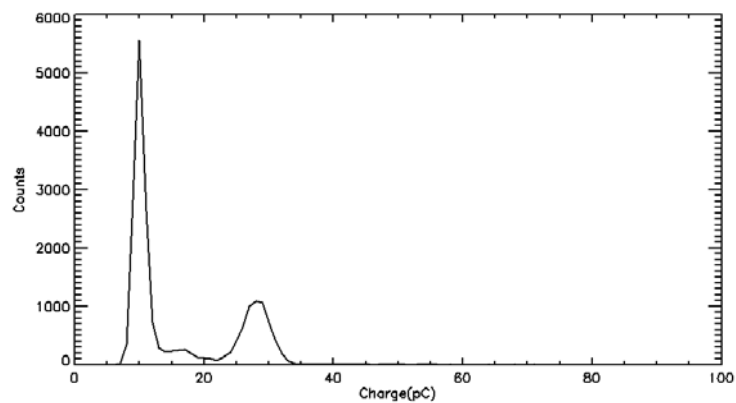

Figure 9: Histogram of sum of charge collected by all 32 SiPM anodes.

The position of the radioactive source along with the lead plate is varied in X and Y direction using XYZ positioning stage. We have carried out the experiment with 1 mm interval in both X and Y direction to see the position detection of the incident X-rays. In our experiment, after selecting the events from the figure 9 (20 pC to 35 pC), we fit two dimensional Gaussian function to the distribution of charge in 36 SiPM anodes (arranged in 6x6 array) for each selected event. Four corner SiPM anodes, are assigned zero values (not connected) while fitting. We make use of MPFIT2DPEAK function of MPFIT library in IDL for this purpose. We get the X and Y position of each of the registered event from this Gaussian fit. Figure 10 shows scatter plot of positions of events measured at four locations of the source 1 mm apart in X and Y direction.

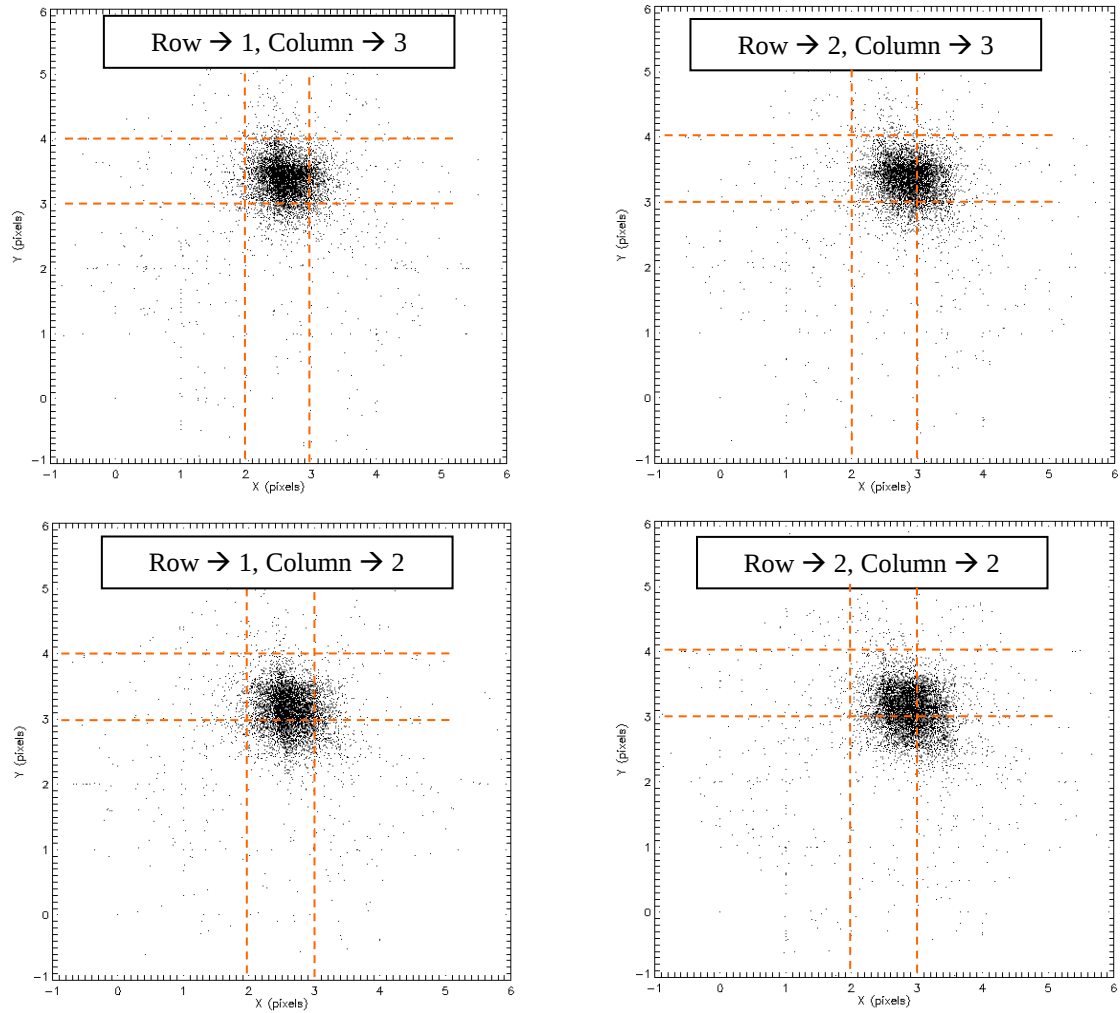

Figure 10: Scatter plot of detected positions of X-ray photons at four locations of X-ray source. See text for details.

One corner of the SiPM is taken as X = 0 mm and Y = 0 mm position. In the figure 10, the radioactive source is moved in both X and Y direction for 1 mm. The diameter of the aperture is ~0.5 mm, and there is ~2 mm gap from $CeBr_3$ top surface to the lead shield and ~1 mm gap between scintillator and SiPM. These gaps add to the dispersion in determined positions of X-ray events. Figure 11 shows the histogram of the X-Y positions for all the events for the same X-ray source positions. The histogram of interaction positions are fitted with two-dimensional Gaussian function to obtain the centroids. X and Y coordinates of the centroid for the four source locations are:

Row 1, Column 3 → 12.65 mm (X), 15.91 mm (Y) (Y)

Row 2, Column 3 → 13.77 mm (X), 15.84 mm (Y)

Row 1, Column 2 → 12.66 mm (X), 14.92 mm (Y) (Y)

Row 2, Column 2 → 13.78 mm (X), 14.8 mm (Y)

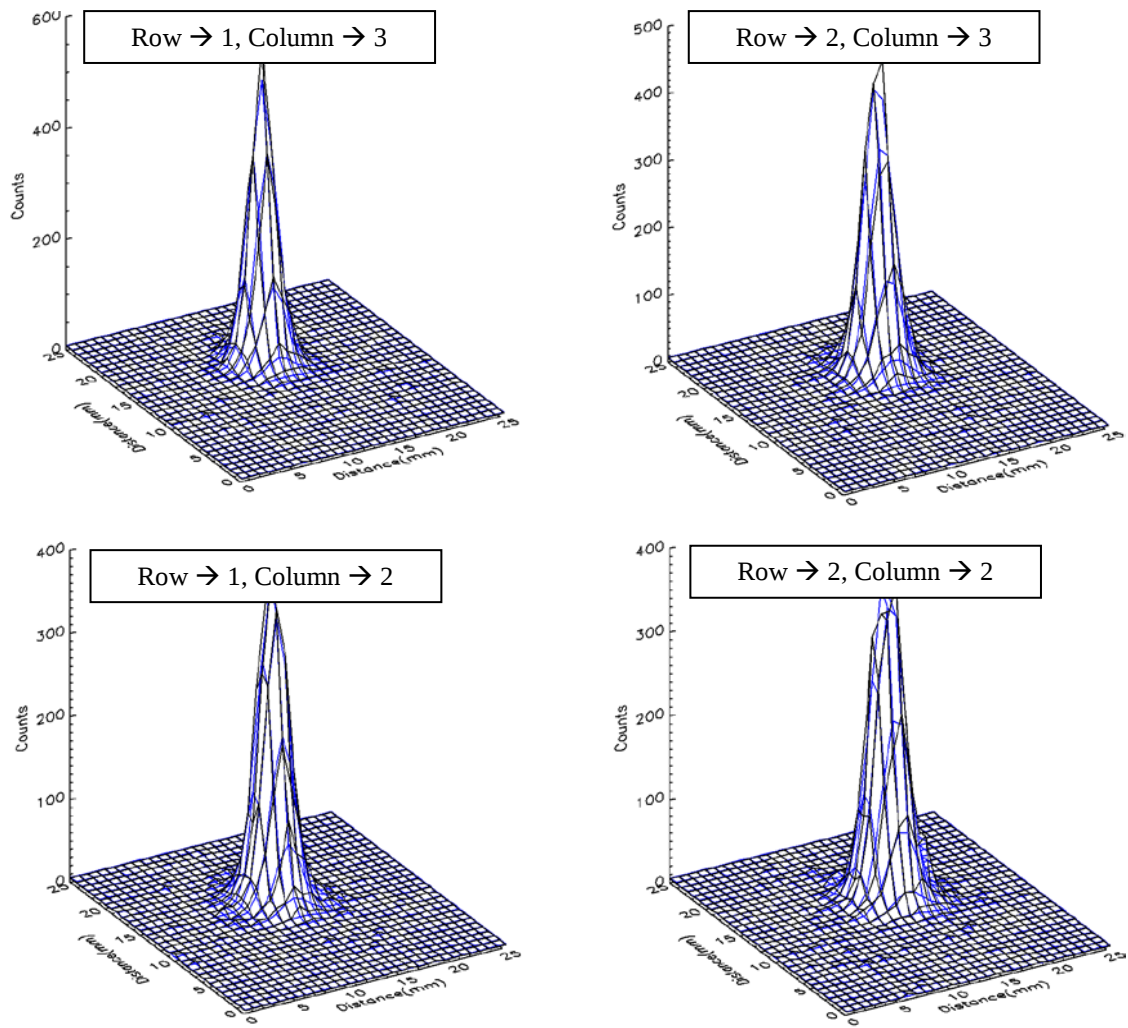

Figure 11: Gaussian fitted histogram of the X and Y positions for all the events for 4 source locations.

### 4.1.2 Results with Vertilon setup for known shape ("T")

In another experiment, a known shape ('T') was made out of the lead. The photographic view of the experiment setup is shown in figure 12. The size of the lead shield used is 25 mm x 25 mm with thickness of 5 mm. $^{241}$Am source was kept at around 15 – 20 cm distance from the lead shield, so that X-rays are passing only through that shape. The processing of the data, acquired from Vertilon remains same i.e. selection of events and then Gaussian fitting to get X-Y position of the events. We performed the experiment with four orientations of the lead shield with respect to scintillator + SiPM array system (0 deg, 45 deg, 90 deg and 180 deg), with approximately $10^5$ events in each case. Results of the experiment are shown in figure 13. The images clearly show the shape of the lead collimator at different orientations.

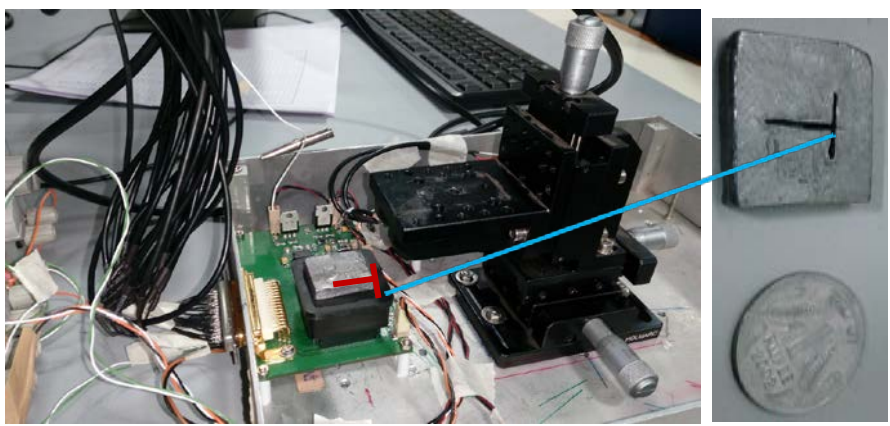

Figure 12: Position imaging for the known shape ("T").

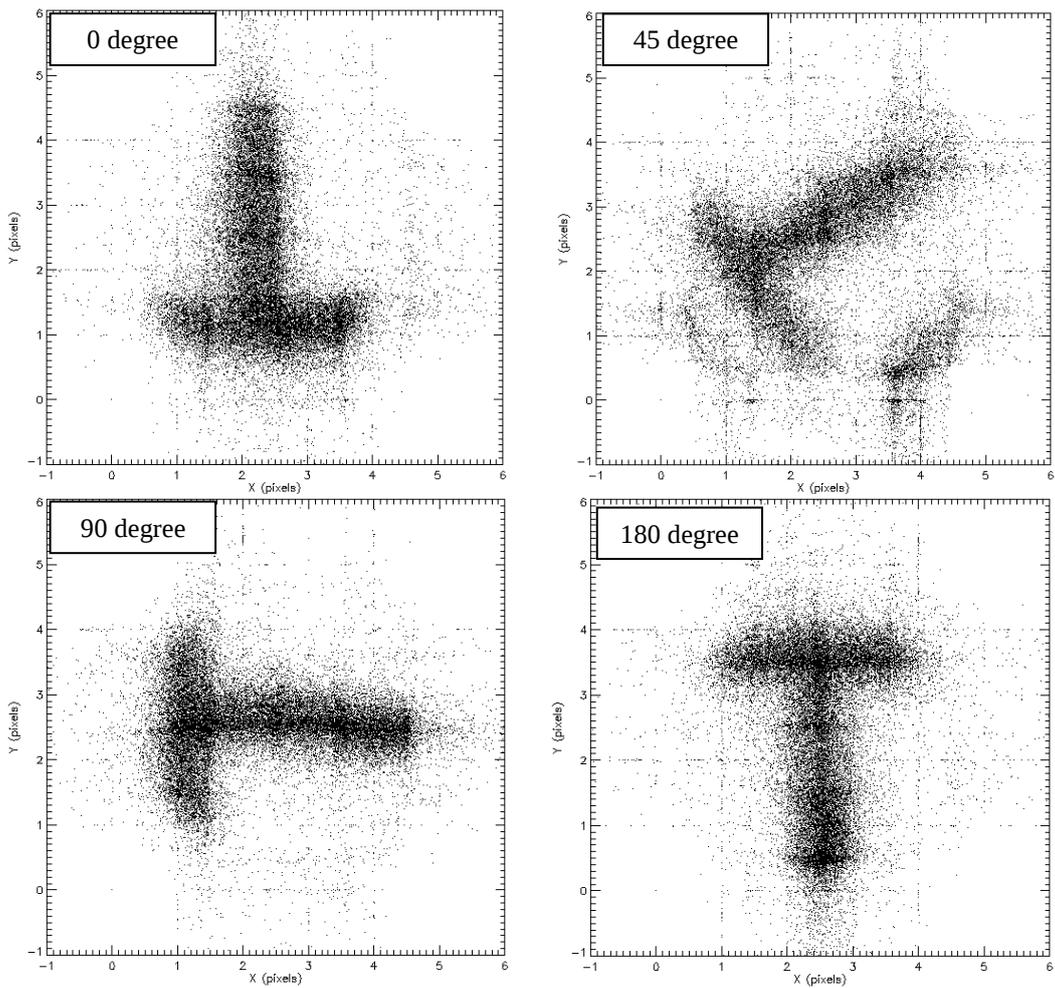

Figure 13: Scatter plot from the known shape ("T").

Position detection using CsI (Tl) scintillator with 4x4 SiPM array is under progress.

### 4.2 Experimental setup using ASIC for hard X-ray position detection

Figure 14 shows the setup using ASIC. In this setup, all 36 SiPM anodes are connected to the IDEAS ASIC board. VATA64HDR16-TG is the ASIC IC, which has readout for 64 channels of SiPM. There are 64 independent analog chains of the CSPA, shaping and peak hold device. This ASIC can be tuned to any of the multiple shaping time constants 50 ns, 100 ns, 150 ns and 300 ns.

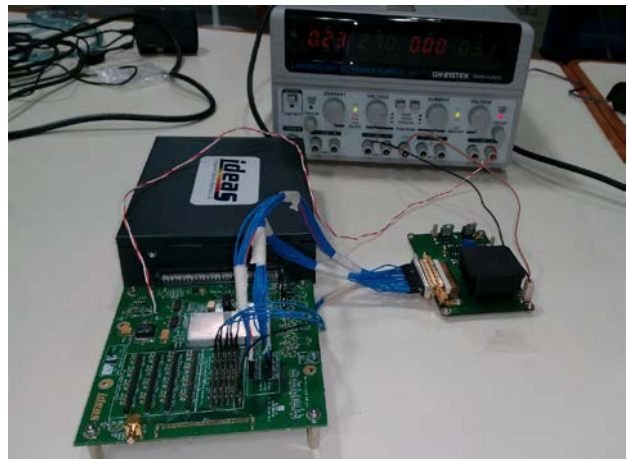

Figure 14: SiPM readout using ASIC.

We are using GALAO development kit for the interfacing between ASIC and the PC. The IDEAS software is being used for the programming of ASIC. In this setup, if anyone of the channels is getting triggered then all the 36 anodes will be readout. The testing of this setup is under progress.

## 5. SUMMARY

We are developing position sensitive detector module using CeBr$_3$ scintillator and SiPM array working in the hard X-ray energy range of 10-400 keV. In this work, we have shown preliminary results of spectral measurement with common cathode readout of SiPM and position measurement with readout of anodes. For spectral measurements with detector module, low energy threshold of ~10 keV is achieved. It is also shown that 1 mm movement of source location can be detected by the position measurements with the detector module. It is to be noted that both spectral and position measurements are expected to improve with optimization of shaping time constants of the readout electronics. This work is under progress using multichannel ASIC for readout of the SiPM array. We also plan to evaluate the performance with use of smaller size SiPM, which will further improve the position resolution.

## 6. ACKNOWLEDGEMENTS

We thank Ms. Vidhi Pandya (PRL), Mr. Shubham (PRL), Ms. Sherry (PRL) and Mr. Arne Fredriksen (IDEAS - ASIC) for their technical support at various stages of this work. This work is supported under Technology Development Program (TDP) of Physical Research Laboratory (PRL), Department of Space, Government of INDIA.